\documentstyle[12pt]{article}
\def\one{1\hskip-.37em 1}
\def\ir{{\rm I}\hskip-.2em{\rm R}}
\def\half{\textstyle{\frac{1}{2}}}

\def\iH{{\rm I}\hskip-.2em{\rm H}}
\def\ra{\rightarrow}
\def\tint{{\textstyle\int}}
\def\d{\partial}
\def\o{\overline}
\def\b{\begin{eqnarray*}}     
\def\e{\end{eqnarray*}}       
\def\bn{\begin{eqnarray}}     
\def\en{\end{eqnarray}}       
\def\<{\langle}
\def\>{\rangle}

\def\{{\lbrace}
\def\}{\rbrace}

\title{Is Quantization Geometry?\footnote{Presented at the ``International 
Conference on 70 Years of Quantum Mechanics and Recent Trends in Theoretical 
Physics'', Calcutta, India, January, 1996.}}
\author{John R. Klauder\\
Departments of Physics and Mathematics\\
University of Florida\\
Gainesville, Fl  32611\\
and\\
I.H.E.S.\\
35 Route de Chartres\\
F-91440 Bures-sur-Yvette}
\begin{document}
\maketitle
\begin{abstract}
The metric known to be relevant for standard quantization procedures 
receives a natural interpretation and its explicit use simultaneously gives 
both physical and mathematical meaning to a (coherent-state) phase-space 
path integral, and at the same time establishes a fully satisfactory, 
geometric procedure of quantization.
\end{abstract}
\section*{Introduction and Background}
\subsubsection*{Purpose and achievements}
The goal of the present work is to motivate and present a {\it conceptually 
simple, fully geometric prescription for quantization}. Such a goal has 
been, and continues to be, the subject of a number of research efforts. In 
the present paper we shall examine the issues from the point of view of 
coherent states. It will be our conclusion that the symplectic geometry of 
classical mechanics, augmented by a natural, and even necessary, metric on 
the classical phase space, are the only essential ingredients to provide a 
quantization scheme that is fully geometric in character and one that can be 
expressed in a coordinate-free form. Moreover, and in a surprising sense, we 
shall see that coherent states---far from being optional---are in fact an 
automatic consequence of this very process of quantization. We shall succeed 
in correctly quantizing a large number (a dense set) of Hamiltonians. In 
that sense the procedures to be described offer a fully satisfactory 
solution to the problem of providing an intrinsic and coordinate-free 
prescription for quantization which agrees with known quantum mechanical 
results. 

Let us begin with a brief overview of just what it is about classical and 
quantum mechanics that is similar, as well as what is different. For ease of 
exposition we shall assume that our phase space essentially admits global 
coordinates. For pedagogical reasons we shall focus on just a single degree 
of freedom.
\subsubsection*{Symplectic manifolds and classical mechanics}
For a single degree of freedom, phase space is a two-dimensional manifold 
$M$ equipped with a {\it symplectic form} $\omega$, which is a closed, 
nondegenerate two form. As a closed form it is locally exact, that is 
$\omega=d\theta$, where $\theta$ is a one form and $d$ denotes the operation 
of exterior derivative. This one form is not unique and it follows from the 
fact that $d^2=0$ that we can replace $\theta$ by $\theta+dG$, where  $G$ is 
a zero form or scalar function. For a general $G$ it follows that
  $$d(\theta+dG)=d\theta+d^2G=d\theta=\omega\;.$$
In addition we introduce a scalar function $H$ on the manifold $M$ which 
takes on a unique value at each point $x\in M$. Now consider a path $x(t)$, 
$0\equiv t'\leq t\leq t''\equiv T$, in the manifold $M$ parameterized by the 
variable $t$ (``time''), so that $\theta$, $G$, and $H$ all become functions 
of $t$ through their dependence on the phase-space points along the path. 
Thus we can construct an action integral according to
  $$    I=\int(\theta+dG-H\,dt)=\int
(\theta-H\,dt)+G''-G'\;,$$
and extremal variations, which hold both end points $x(T)$ and $x(0)$ fixed, 
lead to the sought for classical trajectory through the manifold. The 
associated equations of motion do not involve $G$. In that sense there is an 
{\it equivalence class} of actions all of which lead to the same classical 
equations of motion and thus the same classical trajectory.

We may offer a more familiar formulation with the introduction of canonical 
coordinates. The Darboux theorem assures us that charts of local canonical 
coordinates, say $p$ and $q$, exist for which $\theta=p\,dq$, 
$\omega=dp\wedge dq$, $G=G(p,q)$, and $H=H(p,q)$. In these terms
$$I=\int[p\,dq+dG(p,q)-H(p,q)\,dt]=\int[p\,dq-H(p,q)\,dt]+G(p'',q'')-
G(p',q')\,,$$
where $p'',q''\equiv p(T),q(T)$ and $p',q'\equiv p(0),q(0)$. Insisting that 
the vanishing of the first-order variation holding the end points $p'',q''$ 
and $p',q'$ fixed characterizes the classical trajectory and leads to the 
usual Hamiltonian equations of motion, namely
$${\dot q}=\frac{\d H(p,q)}{\d p}\;,\;\;\;\;\;\;{\dot p}=-
\frac{\d H(p,q)}{\d q}\;.$$
These equations of motion are independent of $G$, and thus we have an 
equivalence class of actions all of which lead to the same classical 
equations of motion.\footnote{We have deliberately derived the Hamiltonian 
equations of motion holding {\it both} $p$ and $q$ fixed at {\it both} the 
initial and the final times. This is normally an {\it over}specification of 
the data, and there will in general be {\it no} solution to the equations of 
motion save for the lucky situation that $p'',q''$ just happens to lie on 
the future evolution of the initial phase space point $p',q'$. Even if we 
overspecify the end point data, we certainly know how to select those 
subcases that will generally admit solutions by focussing on the unique 
solution that follows from choosing just $p',q'$. We are willing to pay the 
price of having no solution, in general, in order that the action principle 
will be invariant under the addition of an arbitrary {\it phase-space} 
function $G(p,q)$ in the integrand; there is no alternative if the action 
principle is to be invariant under canonical coordinate transformations. For 
example, if $G(p,q)=-\half pq$, it is clearly necessary to hold the 
variations of both $p$ {\it and} $q$ fixed at both boundaries in order to 
derive Hamilton's equations of motion. 

There is some ambiguity on this point in the literature \cite{mtw}, which is 
no doubt prompted  by the fact that the phase-space path integral, as it is 
usually formulated (in the Schr\"odinger representation), holds just two end 
point values fixed, namely $q''$ and $q'$, rather than the four end point 
values necessary to achieve full invariance of the action principle under 
general canonical coordinate transformations. As we shall see the 
coherent-state form of the path integral---to which we will automatically be 
led---does indeed specify the values of $p'',q''$ and $p',q'$ consistent 
with the present discussion. Incidentally, in the coherent state case, the 
semi-classical limit leads, in general, to a {\it complex} solution 
\cite{kl2} of the classical equations of motion 
that connect an arbitrary starting pair $p',q'$ with a general final pair 
$p'',q''$.}

New canonical coordinates, say $\o p$ and $\o q$, are invariably related to 
the original canonical coordinates through the one form
  $${\o p}\,d{\o q}=p\,dq+dF({\o q},q)$$
for some function $F$.
It follows, therefore, that for a general $G$ and $F$, one may always find a 
function $\o G$ such that
  $$I=\tint[{\o p}\,d{\o q}+d{\o G}({\o p},{\o q})-{\o H}({\o p},
{\o q})\,dt]=\tint[{\o p}\,d{\o q}-{\o H}({\o p},{\o q})\,dt]+{\o G}(
{\o p''},{\o q''})-{\o G}({\o p'},{\o q'})\,,$$
where ${\o H}({\o p},{\o q})=H(p,q)$.
Extremal variation of this version of the action holding the end points 
fixed as before leads to Hamilton's equations expressed in the form
  $${\dot{\o q}}=\frac{\d{\o H}({\o p},{\o q})}{\d 
{\o p}}\;,\;\;\;\;\;\;{\dot{\o p}}=-\frac{\d{\o H}({\o p},{\o q})}
{\d {\o q}}\;.$$
This form invariance of Hamilton's equations among general canonical 
coordinates is what distinguishes this family of coordinate systems in the 
first place. Such form invariance is the clue that there is, after all, 
something of a geometrical nature underlying this structure, and that 
geometry is in fact the symplectic geometry briefly discussed above.

It is noteworthy that a given Hamiltonian expressed, say, by 
$H(p,q)=\frac{1}{2}(p^2+q^2)+q^4$, in one set of canonical coordinates, 
could, in a new set of canonical coordinates, be expressed simply as $
{\o H}({\o p},{\o q})={\o p}$. If we interpret the first expression as 
corresponding physically to an anharmonic oscillator, a natural question 
that arises is how is one to ``read'' out of the second expression that one 
is dealing with an anharmonic oscillator. This important question will 
figure significantly in our study! 
\subsubsection*{Old quantum theory}
In the old quantum theory one approximately quantized energy levels by the 
Bohr-Sommerfeld quantization scheme in which 
  $$\oint p\,dq=(n+{\half})2\pi\hbar\;,$$
where the integral corresponds to a closed contour in phase space at a 
constant energy value. This integral is to be taken in some canonical 
coordinate system, but which one? Since two canonical coordinate systems are 
related by 
  $${\o p}\,d{\o q}=p\,dq+dF({\o q},q)\;,$$
it follows, when $M$ is simply connected, that 
  $$\oint{\o p}\,d{\o q}=\oint p\,dq\;,$$
and so which canonical coordinates are used doesn't matter---they all give 
the same result. In coordinate-free language this remark holds simply 
because
  $$\oint p\,dq=\int dp\wedge dq=\int d{\o p}\wedge 
d{\o q}\equiv\int\omega\;.$$
Thus we also learn one answer to the question posed above, namely the 
physics of the mathematical expression for the Hamiltonian given simply by 
$\o p$ is coded into the orbits and into the coordinate-invariant 
phase-space areas captured by $\int\omega$ for each value of the energy.
\subsubsection*{New quantum theory}
The royal route to quantization, following Schr\"odinger for example, 
consists of first introducing a Hilbert space of functions $\psi(x)$, 
defined for $x\in\ir$, each of which satisfies $\int|\psi(x)|^2\,dx<\infty$. 
Next the classical phase-space variables $p$ and $q$ are ``promoted'' to 
operators, $p\ra -i\hbar\d/\d x$ and $q\ra x$, which act by differentiation 
and multiplication, respectively. More general quantities, such as the 
Hamiltonian, become operators according to the rule
  $$H(p,q)\ra{\cal H}=H(-i\hbar\d/\d x,x)\;,$$
an expression that may have ordering ambiguities, which we will ignore 
directly, but which in some sense are part of the deeper question: In which 
canonical coordinate systems does such a quantization procedure work? The 
answer of Dirac is: ``This assumption [of replacing classical canonical 
coordinates by corresponding operators] is found in practice to be 
successful only when applied with the dynamical coordinates and momenta 
referring to a Cartesian system of axes and not to more general curvilinear 
coordinates.''\cite{dir}. In other words, the correctness of the 
Schr\"odinger rule of quantization depends on using the right coordinates, 
namely Cartesian coordinates. It is worth emphasizing that Cartesian 
coordinates can only exist on a {\it flat space}. 

An analogous feature is evident in the Feynman phase-space path integral 
formally given by
  $${\cal M}\int e^{(i/\hbar)\int_0^T[p{\dot q}-H(p,q)]\,dt}\,
{\cal D}p\,{\cal D}q\;.$$
 Despite the appearance of this expression as being covariant under a change 
of canonical coordinates, it is, as commonly known, effectively undefined. 
One common way to give meaning to this expression is by means of a 
lattice formulation one manner of which is given by
  $$\lim_{N\ra\infty}M\int\exp\{{\textstyle\frac{i}{\hbar}}\Sigma_0^N[ 
p_{l+\frac{1}{2}}(q_{l+1}-q_l)-\epsilon 
H(p_{l+\frac{1}{2}},{\textstyle\frac{1}{2}}(q_{l+1}+q_l))]\}\,\Pi_0^N\,
dp_{l+\frac{1}{2}}\,\Pi_1^N\,dq_l$$
where $\epsilon\equiv T/(N+1)$, $q_{N+1}\equiv q''$, $q_0\equiv q'$, and 
$M=(2\pi\hbar)^{-(N+1)}$ is a suitable normalization factor. The indicated 
limit is known to exist for a large class of Hamiltonians leading to 
perfectly acceptable (Weyl-ordered) quantizations. However, it is clear that 
unlike the formal continuum path integral, this lattice formulation is {\it 
not} covariant under canonical coordinate transformations; in other words, 
this lattice expression will lead to the correct quantum mechanics only in a 
limited set of canonical coordinates, namely the Cartesian set referred to 
by Dirac. 

This then is the dilemma that confronts us. The ``new'' quantization of 
Schr\"odinger, Heisenberg, and Feynman---the {\it correct} quantization from 
all experimental evidence---seems to depend on the choice of coordinates. 
This is clearly an unsettling state of affairs since nothing physical, like 
quantization, should depend on something so arbitrary as the choice of 
coordinates!
\subsubsection*{Geometric quantization}
There are two attitudes that may be taken toward this apparent dependence of 
the very act of quantization on the choice of coordinates. The first view 
would be to acknowledge the ``Cartesian character'' that is seemingly part 
of the procedure. The second view would be to regard it as provisional and 
seeks to find a quantization formulation that eliminates this apparently 
unphysical feature of the current approaches. 

The goal of eliminating the dependence on Cartesian coordinates in the 
standard approaches is no doubt one of the motivations for several programs 
such as geometric quantization \cite{syn}, deformation quantization 
\cite{fla}, etc. In the first of these programs, for example, one finds the 
basic ingredients: (i) prequantization and (ii) polarization (real and 
complex), which define the framework, i.e, the {\it kinematics}, and (iii) 
one of several proposals to deal with {\it dynamics}. Despite noble efforts, 
it is not unfair to say that to date only a very limited class of dynamical 
systems can be treated in the geometric quantization program which also 
conform with the results of quantum mechanics. 

The approach that we shall adopt takes the other point of view seriously, 
namely that the ``Cartesian character'' is not to be ignored. As we shall 
see when this feature is properly understood and incorporated, {\it a 
genuine geometric interpretation of quantization can be rigorously developed 
that agrees with the standard quantum mechanical results for a wide (dense) 
set of Hamiltonians}. 
\section*{Coherent States}
The concept of coherent states is sufficiently broad by now that there are 
several definitions. Our definition is in reality a rather old one 
\cite{k63}, and a very general one at that, so general that it incorporates 
essentially all other definitions. We start with a label space ${\cal L}$, 
which may often be identified with the classical phase space $M\!$, and a 
continuous map from points in the label space to (nonzero) vectors in a 
Hilbert space $\iH$ (see below). For concreteness, let us choose the label 
space as the phase space for a single degree of freedom. Then each point in 
$M$ may be labelled by canonical coordinates $(p,q)$, and we use that very 
pair to identify the coherent state vector itself: $|p,q\>$ or $\Phi[p,q]$. 
If we choose a different set of canonical coordinates, say $({\o p},{\o q})$ 
to identify the {\it same} point in $M$, then we associate the new 
coordinates to the {\it same} vector $|{\o p},{\o q}\>\equiv|p,q\>$, or even 
better ${\o\Phi}[{\o p},{\o q}]\equiv\Phi[p,q]$.  Although it is by no means 
necessary, we shall specialize to coherent states that are {\it unit 
vectors}, $\<p,q|p,q\>=1=(\Phi[p,q],\Phi[p,q])$, for all points $(p,q)\in 
M$. We place only two requirements on this map from $M$ into $\iH$: 

(1) {\it continuity}, which can be stated as joint continuity in both 
arguments of the coherent state overlap ${\cal 
K}(p'',q'';p',q')\equiv\<p'',q''|p',q'\>$; and 

(2) {\it resolution of unity}, for which a positive measure $\mu$ exists 
such that
  $$\one\equiv\int|p,q\>\<p,q|\,d\mu(p,q)\;,$$
where $\one$ is the unit operator. This last equation may be understood as 
  \b \<\phi|\psi\>\!\!\!&=&\!\!\!\!\int\<\phi|p,q\>\<p,q|\psi\>\,d\mu(p,q)\;,\\
  \<p'',q''|\psi\>\equiv\psi(p'',q'')\!\!\!&=&\!\!\!\!\int{\cal 
K}(p'',q'';p,q)\,\psi(p,q)\,d\mu(p,q)\;,\\
  {\cal K}(p'',q'';p',q')\!\!\!&=&\!\!\!\!\int{\cal K}(p'',q'';p,q)\, {\cal 
K}(p,q;p',q')\,d\mu(p,q)\;.  \e  
Each successive relation has been obtained from the previous one by 
specialization of the vectors involved. The last relation, in conjunction 
with the fact that ${\cal K}(p'',q'';p',q')^*={\cal K}(p',q';p'',q'')$, 
implies that $\cal K$ is the kernel of a {\it projection operator} onto a 
proper subspace of $L^2(\ir^2,d\mu)$ composed of bounded, continuous 
functions that comprise the Hilbert space of interest. The fact that the 
representatives are bounded follows from our choice of coherent states that 
are all unit vectors. Although no group need be involved in our definition 
of coherent states, it is evident that when a group is present 
simplifications may occur. This is true for the canonical coherent states to 
which we now specialize.

With $Q$ and $P$ self adjoint and irreducible, and $[Q,P]=i\hbar$, the 
canonical coherent states defined with help of the unitary Weyl group 
operators for all $(p,q)\in\ir^2\equiv M$ by
$$|p,q\>=e^{-iqP/\hbar}\,e^{ipQ/\hbar}\,|\eta\>\;,\;\;\;\;\;
\<\eta|\eta\>=1\;$$
are a standard example for which $d\mu(p,q)=dp\,dq/2\pi\hbar$. The 
resolution of unity holds in this case for {\it any} normalized fiducial 
vector $|\eta\>$. However, a useful specialization occurs if we insist that 
$(\Omega Q+iP)|\eta\>=0$, $\Omega>0$, leading to the ground state of an 
harmonic oscillator. In that case
$$\<p',q'|p,q\>=\exp\{(i/2\hbar)(p'+p)(q'-q)-(1/4\hbar)[\Omega^{-1}(p'-p)^2+
\Omega(q'-q)^2]\}\;.$$
We note in passing that two more familiar resolutions of unity may be 
obtained as limits. In particular,  \b 
\int\lim_{\Omega\ra\infty}|p,q\>\<p,q|\,dp\,dq/2\pi\hbar=\int|q\>\<q|\,dq=
\one\;,\\ \int\lim_{\Omega\ra 
0}|p,q\>\<p,q|\,dp\,dq/2\pi\hbar=\int|p\>\<p|\,dp=\one\;, \e  where the 
formal vectors $|q\>$ satisfy $Q|q\>=q|q\>$and $\<q'|q\>=\delta(q'-q)$, and 
correspondingly for $|p\>$.

The normalized canonical coherent states $|p,q\>$ that follow from the 
condition $(\Omega Q+iP)|\eta\>=0$ are in fact analytic functions of the 
combination $\Omega q+ip$ apart from a (normalizing) prefactor. When that 
prefactor is removed from the vectors and put into the integration measure, 
one is led directly to the Segal-Bargmann Hilbert space representation by 
holomorphic functions.
\subsubsection*{Symbols}
Generally, and with the notation $\<(\cdot)\>\equiv\<\eta|(\cdot)|\eta\>$, 
it follows from the commutation relations that $\<p,q|P|p,q\>=p+\<P\>$ and 
$\<p,q|Q|p,q\>=q+\<Q\>$; if 
$\<p,q|P|p,q\>=p$ and $\<p,q|Q|p,q\>=q$ we say that the fiducial vector is 
{\it physically centered}. Observe that the labels of the coherent state 
vectors are {\it not eigenvalues} but {\it expectation values}; thus there 
is no contradiction in specifying both $p$ and $q$ simultaneously.   

For a general operator ${\cal H}(P,Q)$ we introduce the {\it upper} symbol
 \b H(p,q)\!\!\!&\equiv&\!\!\!\<p,q|{\cal H}(P,Q)|p,q\>\\ 
&=&\!\!\!\<\eta|{\cal H}(P+p,Q+q)|\eta\>\\
   &=&\!\!\!{\cal H}(p,q)+{\cal O}(\hbar;p,q)\;,  \e
and, when it exists, the {\it lower} symbol $h(p,q)$ implicitly defined 
through the relation
  $${\cal H}=\int h(p,q)\,|p,q\>\<p,q|\,dp\,dq/2\pi\hbar\;. $$
We note that for the harmonic oscillator fiducial vector lower symbols exist 
for a dense set of operators, and generally $H(p,q)-h(p,q)\simeq O(\hbar)$. 
The association of an operator $\cal H$ with the function $h(p,q)$ is an 
example of what goes under the name of Toeplitz quantization today 
\cite{jaf}. 
\subsubsection*{Differentials}
Several differential expressions are already implicitly contained within the 
coherent states. The first is the canonical one form
  $$\theta\equiv 
i\hbar\<\;|d|\;\>=i\hbar(\Phi,d\Phi)=i\hbar\Sigma_n\phi^*_n\,d\phi_n$$
written in coordinate-free notation, or alternatively,
  $$\theta=i\hbar\<p,q|d|p,q\>=p\,dq+\<P\>\,dq-\<Q\>\,dp=p\,dq $$
using canonical coordinates, and where we have ended with a physically 
centered fiducial vector. In coordinate-free notation it follows that
  $$\omega\equiv d\theta =i\hbar\Sigma_n d\phi^*_n\wedge d\phi_n\;,$$
and in canonical coordinates that
  $$\omega=dp\wedge dq=d{\o p}\wedge d{\o q}\;,$$
along with $d\omega=0$ which follows directly. A useful Riemannian metric 
is given first in coordinate-free notation by
  \b  d\sigma^2\!\!\!&\equiv&\!\!\! 2\hbar^2
[ |\!|d|\;\>|\!|^2-|\<\;|d|\;\>|^2]\\    
&=&\!\!\!2\hbar^2\Sigma_{n,m}\,d\phi^*_n(\delta_{nm}-\phi_n\phi^*_m)
d\phi_m\;,   \e
and second in canonical coordinates by
  \b d\sigma^2(p,q)\!\!\!&=&\!\!\! \hbar(dp^2+dq^2)\;,\;\;\;\;\;\;\;\;\; 
(\Omega=1)\;,\\
  d\sigma^2({\o p},{\o q})\!\!\!&=&\!\!\!\hbar[A({\o p},{\o q})
d{\o p}^2+B({\o p},{\o q})d{\o p}\,d{\o q}+C({\o p},{\o q})d{\o q}^2]\;.  \e
In the next to the last line the line element is expressed in the Cartesian 
form it takes for a Gaussian fiducial vector, while in the last line is the 
expression of the flat metric in general canonical coordinates.
\subsubsection*{Canonical and unitary transformations}
In classical mechanics canonical transformations may either be viewed as 
passive or active. Passive transformations leave the point in phase space 
fixed but change the coordinates by which it is described; active 
transformations describe a flow of points in phase space against a fixed 
coordinate system. Perhaps the best known example of an active 
transformation is the continuous unfolding in time of a dynamical evolution. 
In quantum mechanics unitary transformations are presumed to play the role 
that canonical transformations play in the classical theory \cite{shi}. If 
$p\ra P$ and $q\ra Q$, then it follows that 
${\o p}=(p+q)/\sqrt{2}\ra(P+Q)/\sqrt{2}\equiv{\o P}$ and 
${\o q}=(q-p)/\sqrt{2}\ra(Q-P)/\sqrt{2}\equiv{\o Q}$, and moreover there 
exists a unitary operator $U$ such that ${\o P}=U^\dagger PU$ and 
${\o Q}=U^\dagger QU$. Consider instead the classical canonical 
transformation ${\tilde p}\equiv(p^2+q^2)/2\ra{\tilde P}$ and ${\tilde 
q}\equiv\tan^{-1}(q/p)\ra{\tilde Q}$. As basically a transformation to polar 
coordinates this canonical transformation is well defined except at the 
single point $p=q=0$. However, the associated quantum operators in this case 
cannot be connected by a unitary transformation to the original operators 
$P$ and $Q$ (because ${\tilde P}\geq0$ and the spectrum of an operator is 
preserved under a unitary transformation). Thus some passive canonical 
transformations have images in unitary transformations while others 
definitely do not.  

Using coherent states it is possible to completely disconnect canonical 
transformations and unitary transformations. Consider the transformations of 
the upper and lower symbols in the following example:
  \b {\half}(p^2+q^2)\!\!\!&=&\!\!\!\<p,q|\,{\half}(P^2+Q^2-\hbar)|p,q\>\\&=&
\!\!\!\<{\tilde p},{\tilde q}|\,{\half}(P^2+Q^2-\hbar)|{\tilde p},
{\tilde q}\>={\tilde p}\;,\\ 
{\half}(P^2+Q^2+\hbar)\!\!\!&=&\!\!\!\int{\half}(p^2+q^2)\,|p,q\>\<p,q|\,dp
\,dq/2\pi\hbar\\&=&\!\!\!\int{\tilde p}\,|{\tilde p},{\tilde q}\>
\<{\tilde p},{\tilde q}|\,d{\tilde p}\,d{\tilde q}/2\pi\hbar\;.  \e
Observe in this example how the {\it operators} and {\it coherent state 
vectors} have remained {\it completely fixed} as the coordinates have passed 
from $(p,q)$ to $({\tilde p},{\tilde q})$. Of course, one may also introduce 
separate and arbitrary unitary transformations of the operators and vectors, 
e.g. $P\ra VPV^\dagger$, and $|p,q\>\ra V|p,q\>$, etc., which have the 
property of preserving inner products.
\section*{The Shadow Metric} 
\subsubsection*{A Flat Metric is Already Present}
The form of the metric $d\sigma^2$ was given earlier for a special (harmonic 
oscillator) fiducial vector. If instead we consider a {\it general fiducial 
vector} $|\eta\>$, then it follows that
  $$d\sigma^2=\<(\Delta Q)^2\>\,dp^2+\<\Delta P\Delta Q +\Delta Q\Delta P\>\,dp\,dq+\<(\Delta P)^2\>\,dq^2\;,$$
which shows itself to be always {\it flat}; thus this is a property of the 
{\it Weyl group} and not of the fiducial vector. Here $\Delta P\equiv 
P-\<P\>$,  etc. Unlike the symplectic form or the Hamiltonian, for example, 
the metric is typically $O(\hbar)$ and thus it is essentially nonclassical. 

Indeed, any quantization scheme in which the Weyl operators and Hilbert 
space vectors exist leads to the metric $d\sigma^2$, whether it is 
intentional or not. Such quantization schemes may not explicitly {\it use} 
the metric, but it is nevertheless {\it there}. Moreover, any proposed 
quantization scheme that vigorously avoids introducing a metric must come to 
terms with how a flat space appears, as it must, if such a quantization 
scheme ultimately involves Weyl operators and Hilbert space vectors.
\subsubsection*{Cartesian Coordinates are Important}
We assert that physics actually resides in Cartesian coordinates, and more 
generally in the coordinate form of the metric. Suppose 
$d\sigma^2=\hbar(dp^2+dq^2)$, then it follows that 
$H(p,q)=\frac{1}{2}(p^2+q^2)$ implies, as before, that 
${\cal H}=\frac{1}{2}(P^2+Q^2-\hbar)$. On the other hand, if instead 
$d\sigma^2=\hbar[(2{\tilde p})^{-1}d{\tilde p}^2+(2{\tilde p})d
{\tilde q}^2]$, then ${\tilde H}({\tilde p},{\tilde q})={\tilde p}$ implies 
that ${\cal H}=\frac{1}{2}(P^2+Q^2-\hbar)$. In other words, {\it the 
physical meaning of the coordinatized mathematical expression for some 
classical quantity is coded into the coordinate form of the metric!} This 
remark is already true at the classical level, namely one needs a flat 
``shadow'' metric (sometimes called a ``secret'' metric) on the classical 
phase space, or at least on a copy of it, so that one can ascribe physical 
meaning to the coordinatized mathematical expressions for one or another 
classical quantity. If the flat shadow metric is expressed in Cartesian 
coordinates, then one may interpret an expression such as 
$\frac{1}{2}(p^2+q^2)+q^4$ as truly representing a physical, quartic 
anharmonic oscillator; if the flat shadow metric is {\it not} expressed in 
Cartesian coordinates, then no such physical interpretation of such a  
mathematical expression is justified. 

Although we have originally introduced the phase-space metric in the quantum 
theory via its construction in terms of coherent states, we now see that we 
can alternatively view the phase-space metric (modulo a coefficient $\hbar$) 
as an auxiliary classical expression that provides physical meaning for 
coordinatized expressions of the classical theory.
\section*{Quantization and Continuous-time \\Regularization}
It should be self evident that quantization relates to physical systems 
inasmuch as the quantization of a particular Hamiltonian is designed to 
generate the spectrum appropriate to that physical system. Consider again 
the formal phase-space path integral given by
  $${\cal M}\int e^{(i/\hbar)\int_0^T[p{\dot q}+
{\dot G}(p,q)-h(p,q)]\,dt}\,{\cal D}p\,{\cal D}q\;.$$
We have already stressed that this expression is not mathematically defined, 
and now we emphasize that in fact it has no physics as well because there is 
no way of telling to which physical system the coordinatized expression for 
the Hamiltonian corresponds. In short, the formal phase-space path integral 
expression has neither mathematical nor physical meaning as it stands! 

We will remedy this situation in a moment, but there is one ``toy'' analog 
worth introducing initially. Consider the conditionally convergent integral 
that is given a definition through the introduction of a regularization and 
its removal as in the expression
  $$\int_{-\infty}^\infty 
e^{iy^2/2}\,dy\equiv\lim_{\nu\ra\infty}\int_{-\infty}^{\infty}
e^{iy^2/2-y^2/2\nu}\,dy=\sqrt{2\pi i}\;.$$
Other regularizations may lead to the same answer, or in fact they may lead 
to different results; the physical situation should be invoked to choose the 
relevant one. 

Now let us introduce a continuous-time convergence factor into the formal 
phase-space path integral in the form 
 \b\lim_{\nu\ra\infty}\!\!\!\!\!\!&&\!\!\!\!\!\!{\cal M}_\nu\int 
e^{(i/\hbar)\int_0^T[p{\dot q}+
{\dot G}(p,q)-h(p,q)]\,dt}\,e^{-(1/2\nu)\int_0^T({\dot p}^2+
{\dot q}^2)\,dt}\,{\cal D}p\,{\cal D}q\\
 &=&\!\!\lim_{\nu\ra\infty}2\pi\hbar\,e^{\hbar\nu T/2}\int 
e^{(i/\hbar)\int_0^T[p\,dq+dG(p,q)-h(p,q)\,dt]}\,d\mu_W^\nu(p,q)\;.  \e
In the first line we have formally stated the form of the regularization, 
while in the second line appears the proper mathematical statement it 
assumes after some minor rearrangement. The measure $\mu_W^\nu$ is a 
two-dimensional Wiener measure expressed in Cartesian coordinates on the 
plane as signified by the metric $dp^2+dq^2$ that appears in the first line 
in the regularization factor. {\it Here enters explicitly the very shadow 
metric itself, used to give physical meaning to the coordinatized form of 
the Hamiltonian, and which now additionally underpins a rigorous 
regularization for the path integral!} As Brownian motion paths, with 
diffusion constant $\nu$, almost all paths are continuous but nowhere 
differentiable. Thus the initial term $\int p\,dq$ needs to be defined as a 
{\it stochastic integral} and we choose to do so in the Stratonovich form, 
namely as $\lim \Sigma\frac{1}{2}(p_{l+1}+p_l)(q_{l+1}-q_l)$, where 
$q_l\equiv q(l\epsilon)$, etc., and the limit refers to $\epsilon\ra0$ 
\cite{str}. This prescription is generally different from that of It\^o, 
namely $\lim \Sigma p_l(q_{l+1}-q_l)$, due to the unbounded variation of the 
Wiener paths involved. Observe, in the second line above, for each 
$0<\nu<\infty$, that {\it no mathematical ambiguities remain}, i.e., the 
expression is completely well defined. As we note below not only does the 
limit exist but it also provides the correct solution to the Schr\"odinger 
equation  for a dense set of Hamiltonian operators.

The continuous-time regularization, as well as its Wiener measure 
counterpart, involves pinning the paths $p(t),q(t)$ at $t=T$ and at $t=0$ so 
that $(p'',q'')=(p(T),q(T))$ and $(p',q')=(p(0),q(0))$. This leads to an 
expression of the form $K(p'',q'',T;p',q',0)$, which may be shown to be
  \b K(p'',q'',T;p',q',0)\!\!\!&\equiv&\!\!\!\<p'',q''|\,
e^{-i{\cal H}T/\hbar}|p',q'\>\;,\\
    |p,q\>\!\!\!&\equiv&\!\!\! 
e^{-iG(p,q)/\hbar}\,e^{-iqP/\hbar}\,e^{ipQ/\hbar}\,|\eta\>\;,\;\;\;\;
(Q+iP)|\eta\>=0\;,\\
   {\cal H}\!\!\!&\equiv&\!\!\!\int h(p,q)|p,q\>\<p,q|\,dp\,dq/2\pi\hbar\;. 
 \e
In brief, the regularization chosen {\it automatically} leads to a coherent 
state representation, and, in addition, it {\it selects} the Hamiltonian 
operator determined by the lower symbol. There are three technical 
requirements for this representation to hold \cite{dkl}: 
\b  (1)&&\;\;\;\;\;\int h^2(p,q)\,e^{-A(p^2+q^2)}\,dp\,dq<\infty\;,\;\;\;\; 
{\rm for \; all}\;\;A>0\;,\\
   (2)&&\;\;\;\;\;\int h^4(p,q)\,e^{-B(p^2+q^2)}\,dp\,dq<\infty\;,\;\;\;\; 
{\rm for\; some}\;\; B<{\half}\;,\\
  (3)&&\;\;\;\;\;{\cal H}\;\; {\rm \;is\; e. s. a.\; on }
\;\;D=\{\Sigma_0^Na_n|n\>:\;a_n\in{\bf C},\;N<\infty\}\;,  \e
where the orthonormal states 
$|n\>\equiv(1/\sqrt{n!})[(Q-iP)/\sqrt{2\hbar}]^n|\eta\>$, $n\geq0$.
Thus this representation includes (but is not limited to) {\it all 
Hamiltonians that are Hermitian, semibounded polynomials of the basic 
operators $P$ and $Q$}. We note that $G$ generally serves as an unimportant 
gauge; however, if the topology of $M$ is not simply connected then $G$ 
contains the Aharanov-Bohm phase \cite{ree}. Observe that the propagator 
formula also has an {\it analog physical system}, namely a two-dimensional 
particle moving on a flat plane in the presence of a constant magnetic field 
perpendicular to the plane. The limit in which the mass of the particle goes 
to zero projects the system onto the first Landau level.

The point of using the Stratonovich prescription for stochastic integrals is 
that the ordinary rules of calculus still apply \cite{str}. Thus the rule 
for a canonical transformation given earlier, namely ${\o p}\,
d{\o q}=p\,dq+dF({\o q},q)$, still applies to Brownian motion paths. 
Consequently, just as in the classical case a function ${\o G}({\o p},
{\o q})$ exists so that after such a canonical coordinate transformation
 \b {\o K}({\o p}'',{\o q}'',T;{\o p}',{\o q}',0)\!\!\!\!&=&\!\!\!\!
\<{\o p}'',{\o q}''|\,e^{-i{\cal H}T/\hbar}\,|{\o p}',{\o q}'\>\\
  \!\!\!\!&=&\!\!\!\!\lim_{\nu\ra\infty}2\pi\hbar\, e^{\hbar\nu T/2}\int 
e^{(i/\hbar)\int_0^T[{\o p}d{\o q}+d{\o G}({\o p},{\o q})-{\o h}({\o p},
{\o q})dt]}\,d{\o\mu}_W^\nu({\o p},{\o q})\;. \e
Here ${\o\mu}_W^\nu$ denotes two-dimensional Wiener measure on the flat 
plane expressed in general canonical coordinates. 
\subsubsection*{Coordinate-free formulation}
The covariant transformation of the propagator indicated above implies that 
a coordinate-free representation exists. We first introduce Brownian motion 
as a map $\rho(t;0):\,M\times M\ra\ir^+$, $t>0$, with 
$\lim_{t\ra0}\rho=\delta$, $\d\rho/\d t=(\nu/2)\Delta\rho$, and finally 
$\rho(t;0)=\int d\mu_W^\nu$ which defines a coordinate-free Wiener measure. 
Next we let $\phi:\,M\ra\bf C$, ${\cal K}:\,M\times M\ra\bf C$, 
$\phi\in({\cal K})L^2(M,\omega)$, and $\phi={\cal K}\phi$ (N.B. $\cal K$ is 
an analog of ``polarization''). Quantum dynamics comes from $i\hbar\d\phi/
\d t={\cal H}\phi$, where ${\cal H}={\cal K}h{\cal K}$ (N.B. this relation 
has the effect of ``preserving polarization''); also we introduce 
$K(T;0):\,M\times M\ra\bf C$, so that $\phi(T)=K(T;0)\phi(0)$. The 
construction of the reproducing kernel $\cal K$ and the propagator $K$ reads
  \b  {\cal K}\!\!&\equiv&\!\!\lim_{\nu\ra\infty}2\pi\hbar\, e^{\hbar\nu 
T/2}\int e^{(i/\hbar)\int(\theta+dG)}\,d\mu_W^\nu=\lim_{T\ra0}K(T;0)\;,\\
  K(T;0)\!\!&\equiv&\!\!\lim_{\nu\ra\infty}2\pi\hbar\, e^{\hbar\nu T/2}\int 
e^{(i/\hbar)\int(\theta+dG-h\,dt)}\,d\mu_W^\nu\;.  \e
Observe again how a flat metric has been used for the Brownian motion; in 
our view it is this flat (phase) space that underlies Dirac's remark related 
to canonical quantization quoted above.
\subsubsection*{Alternative continuous-time regularizations}
Our introduction of Brownian motion on a {\it flat} two-dimensional phase 
space has led to canonical quantization, namely one involving the Heisenberg 
operators $P$ and $Q$. If instead we choose to regularize on a phase space 
taken as a two-dimensional {\it spherical} surface of radius $R$, where 
$R^2\equiv s=\hbar/2,\;\hbar,\;3\hbar/2,\ldots$, then such a Brownian motion 
regularization automatically leads to a quantization in which the 
kinematical operators are the spin operators $S_1,\;S_2,$ and $S_3$ such 
that $\Sigma S_j^2=s(s+1)\hbar^2$, i.e., the generators of the SU(2) group 
\cite{dkl}. In like manner, if we introduce a Brownian motion regularization 
on a two-dimensional {\it pseudo-sphere} of constant negative curvature, 
then the kinematical operators that automatically emerge are the generators 
of the affine (``$ax+b$'') group, a subgroup of SU(1,1)\cite{dkp}. The three 
examples given here exhaust the simply connected spaces of constant 
curvature in two dimensions; they also have the property that the metric 
assumed for the Brownian motion regularization coincides with the metric 
that follows from the so-derived coherent states.

Summarizing, {\it the geometry of the regularization that supports the 
Brownian motion actually determines the nature of the kinematical operators 
in the quantization!}
\section*{Regularization on a General 2-D Surface}
Finally, we note that the present kind of quantization can be extended to a 
general two-dimensional surface without symmetry and with an arbitrary 
number of handles. We only quote the result for the propagator. Let 
$\xi^j,\;j=1,2,$ denote the two coordinates, $g_{jk}(\xi)$ the metric,
$a_j(\xi)$ a two-vector, $f_{jk}(\xi)=\d_ja_k(\xi)-\d_ka_j(\xi)$ its curl, 
and $h(\xi)$ the classical Hamiltonian. Then the propagator is defined by
\cite{akl}
 \b  \<\xi'',T|\xi',0\>\!\!&=&\!\!\<\xi''|\,e^{-i{\cal H}T/\hbar}|\xi'\>\\
&=&\!\!\lim_{\nu\ra\infty}
{\cal M}_\nu\int\exp\{(i/\hbar)\int[a_j(\xi){\dot\xi}^j-h(\xi)]\,dt\}\\
&&\times\exp\{-(1/2\nu)\int g_{jk}(\xi){\dot\xi}^j{\dot\xi}^k\,dt\}\\
&&\times\exp\{(\hbar\nu/4)\int\sqrt{g(\xi)}\epsilon^{jk}f_{jk}(\xi)\,dt\}
\,\Pi_t\sqrt{g(\xi)}\,d\xi^1\,d\xi^2\;.  \e
Observe, in this general setting, that the phase-space metric tensor 
$g_{jk}(\xi)$ is one of the necessary {\it inputs} to the process of 
quantization under discussion. From this viewpoint the phase-space metric 
induced by the coherent states is regarded as a derived quantity, and in the 
general situation the two metrics may well differ.
For a compact manifold $M$ it is necessary that $\int 
f_{jk}(\xi)d\xi^j\wedge d\xi^k=4\pi\hbar n$, $n=1,2,3,\ldots\,$. In this 
case the Hilbert space dimension $D=n+1-{\o g}$, where ${\o g}$ is the 
number of handles in the space (genus). Here the states $|\xi\>$ are 
coherent states that satisfy
\b  \one\!\!&=&\!\!\int|\xi\>\<\xi|\,\sqrt{g(\xi)}\,d\xi^1\,d\xi^2\;,\\
   {\cal H}\!\!&=&\!\!\int 
h(\xi)\,|\xi\>\<\xi|\,\sqrt{g(\xi)}\,d\xi^1\,d\xi^2\;.  \e
Observe that although the states $|\xi\>$ are coherent states in the sense 
of this article, there is generally {\it no transitive group} with which 
they may be defined. The propagator expression above is manifestly covariant 
under arbitrary coordinate transformations, and a gauge transformation of 
the vector $a$ introduces a gauge-like contribution that does not appear in 
the field $f$. Finally---and contrary to general wisdom---we note that the 
weighting in the case of a general geometry is {\it nonuniform} in the sense 
that the symplectic form $\omega=f_{jk}\,d\xi^j\wedge d\xi^k/2$ is generally 
{\it not} proportional to the volume element $\sqrt{g(\xi)}\,d\xi^1\,d\xi^2$ 
needed in the resolution of unity and hence in the path integral 
construction.
\section*{Acknowledgements}
It is a pleasure to acknowledge the organizers, especially P. Bandyopadhyay, 
for their efforts in making the conference a success. Thanks are expressed 
to R. Jackiw for raising several points that resulted in the introduction of 
additional clarifying comments.

\end{document}